\documentclass{article}

     \PassOptionsToPackage{numbers, compress}{natbib}

\usepackage[preprint]{neurips_2023}

\usepackage[utf8]{inputenc} 
\usepackage[T1]{fontenc}    
\usepackage{hyperref}       
\usepackage{url}            
\usepackage{booktabs}       
\usepackage{amsfonts}       
\usepackage{nicefrac}       
\usepackage{microtype}      
\usepackage{xcolor}         
\usepackage{amsmath}
\usepackage{amssymb}
\usepackage{mathtools}
\usepackage{amsthm}
\usepackage{multirow}
\usepackage{wrapfig}
\usepackage[capitalize,noabbrev]{cleveref}
\usepackage{graphicx}
\usepackage{subfigure} 
\usepackage{float}
\usepackage{algorithm} 
\usepackage{algorithmic}
\usepackage{textcomp} 
\usepackage{listings} 
\usepackage{enumitem}
\usepackage{utfsym}
\usepackage{times}
\usepackage{latexsym}
\usepackage{fancyvrb}
\title{UniAudio 1.5: Large Language Model-driven Audio Codec is A Few-shot Audio Task Learner}

\author{Dongchao Yang$^{1}$, Haohan Guo$^{1}$, Yuanyuan Wang$^{2}$, Rongjie Huang$^{1}$, Xiang Li$^{2}$ \\
\textbf{Xu Tan$^{3}$, Xixin Wu$^1$,  Helen Meng$^{1}$}  \\
$^1$ The Chinese University of Hong Kong,
$^2$ Tsinghua University, 
$^3$ Microsoft Research Asia \\ 
\texttt{dcyang@se.cuhk.edu.hk} \\
}

\begin{document}

\maketitle
\begin{abstract}
The Large Language models (LLMs) have demonstrated supreme capabilities in text understanding and generation, but cannot be directly applied to cross-modal tasks without fine-tuning. This paper proposes a cross-modal in-context learning approach, empowering the frozen LLMs to achieve multiple audio tasks in a few-shot style without any parameter update. Specifically, we propose a novel and LLMs-driven audio codec model, LLM-Codec, to transfer the audio modality into the textual space, \textit{i.e.} representing audio tokens with words or sub-words in the vocabulary of LLMs, while keeping high audio reconstruction quality. The key idea is to reduce the modality heterogeneity between text and audio by compressing the audio modality into a well-trained LLMs token space. Thus, the audio representation can be viewed as a new \textit{foreign language}, and LLMs can learn the new \textit{foreign language} with several demonstrations. In experiments, we investigate the performance of the proposed approach across multiple audio understanding and generation tasks, \textit{e.g.} speech emotion classification, audio classification, text-to-speech generation, speech enhancement, etc. The experimental results demonstrate that the LLMs equipped with the proposed LLM-Codec, named as UniAudio 1.5, prompted by only a few examples, can achieve the expected functions in simple scenarios. It validates the feasibility and effectiveness of the proposed cross-modal in-context learning approach. To facilitate research on few-shot audio task learning and multi-modal LLMs, we have open-sourced the LLM-Codec model. \footnote{https://github.com/yangdongchao/LLM-Codec}

\end{abstract}

\section{Introduction}

Large language models (LLMs) (\textit{e.g.} GPT4 \citep{gpt4}, LLAMA \citep{llama}) become more versatile and effective at handling diverse and complex Natural Language Processing (NLP) tasks after scaling their model and training data. It is worth noting that the in-context learning ability of LLMs can be used to solve unseen tasks, \textit{e.g.} we can provide instruction along with a few demonstrations of the new task, and then LLMs learn to solve the task. The success of LLMs inspires us to build multi-modal LLMs to solve audio-related tasks in the audio domain. A natural idea is to empower the auditory sense of the LLMs. There have been notable advancements in extending the capabilities of LLMs to tackle audio understanding tasks by combining the pre-trained audio encoder (\textit{e.g.} Whisper encoder \cite{whisper}) and LLMs. For example, WavLLM \cite{hu2024wavllm}, SALMONN \cite{tang2023salmonn} and Qwen-audio \cite{qwen-audio} propose to train a multi-modal LLMs based on a pre-trained audio encoder, a trainable adaptor, and pre-trained LLMs. They try to align the audio and text modalities by updating the adaptor or fine-tuning the LLMs with LORA \cite{hu2021lora}. However, previous works (1) focus more on expanding LLMs to solve specific audio tasks, without considering the in-context-learning ability to unseen audio tasks; (2) do not support audio generation tasks, which limits its application scenarios; (3) align the audio and text modalities by collecting large-scale audio task data to train the models, which increases the efforts for the model training and data collection.

\begin{figure*}[!t]
    \centering
    \includegraphics[width=\textwidth]{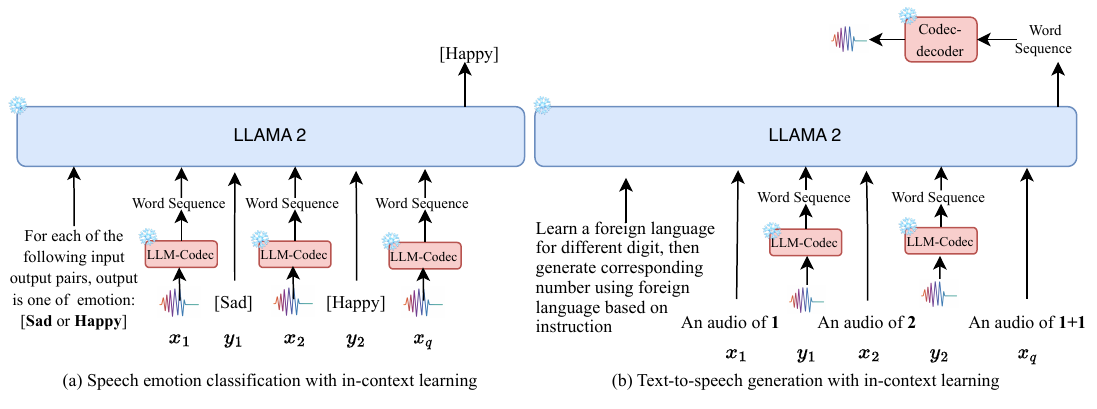}
    \vspace{-0.6cm}
    \caption{This figure illustrates the framework of the proposed
approach (UniAudio 1.5) to conduct speech emotion classification and simple text-to-speech generation tasks. The data format will be $\{x_1,y_1, x_2,y_2, ..., x_q\}$, which means the previous samples $\{ x_i, y_i \}$ is the demonstration of this task, the LLAMA model is asked to predict $y_q$. $y_q$ can be the text or audio.} 
    \label{fig:undersanding}
    \vspace{-0.6cm}
\end{figure*}

In this study, we propose a cross-modal in-context learning approach, empowering the frozen LLMs to solve any user-defined audio tasks based on a few demonstrations without any parameter update. To realize this target, we propose to learn a vector quantization audio codec model to map an audio modality to the token space of a frozen LLMs (\textit{e.g.} LLAMA 2 \cite{llama}), named LLM-Codec. Our motivation is to reduce the modality heterogeneity between audio and text by compressing the audio data into a token space of LLMs. Considering that the compressed audio by LLM-Codec and text modality have a shared vocabulary, the compressed audio sequence can be viewed as a new \textit{foreign language}, and LLMs can learn the new \textit{foreign language} with several demonstration samples. Furthermore, LLMs are pre-trained on large-scale data and discover many patterns of combination of token sequence, which potentially improves its generalization to the \textit{foreign language}. Figure \ref{fig:undersanding} shows how to combine proposed LLM-Codec and LLAMA 2 models for audio tasks.

The proposed LLM-Codec tries to compress the audio data into a lexical word sequence. A desired LLM-Codec should have the following properties: (1) \textbf{Completeness} \cite{10114504}: it should recover the compressed audio without too much loss. (2) \textbf{Compactness}: it should encode the audio into fewer-token sequences. (3) \textbf{Semantic richness}: it should encode the audio into a semantic-rich token sequence, which is easier to recognize by pre-trained LLMs. Thus, we propose a semantic-guided multi-scale residual vector quantization (RVQ) based codec. Specifically, the codec model has three residual VQ layers, the first VQ layer tries to encode the semantic information, the second VQ layer tries to encode the coarse-grained acoustic information, and the third layer tries to encode the residual acoustic information. Different from previous works \cite{encodec,soundstream}, which encode the audio data into the same granularity in each layer, we propose to encode the audio data into different granularity in each layer. Our motivation is that semantic-level information can also be preserved with few tokens. Instead, acoustic-level information needs more tokens. Such a multi-scale setting not only reduces the length of the token sequence but also provides a flexible choice for different types of tasks, \textit{e.g.} for the audio understanding task, we can only use the semantic-level VQ layer. Furthermore, a novel semantic loss and consistency loss are designed to train the LLM-Codec model better. \\
We conduct experiments to verify the effectiveness of LLM-Codec in an in-context learning setting. We use the pre-trained LLAMA 2 7B model for all experiments without any parameter updating. We design audio understanding and generation tasks to evaluate the effectiveness of the proposed method, including speech emotion classification, audio classification, simple text-to-speech, speech denoising, and so on. The main contributions of this work are summarized as follows: \\
\begin{itemize} 
\vspace{-0.7cm}
\item  We propose a novel LLMs-driven audio codec model, LLM-Codec, which effectively connects the text and audio modalities. To the best of our knowledge, this is the first work to quantize the audio data into the representation space of LLMs. 
\vspace{-0.3cm}
\item  We demonstrate the feasibility and potential of using the in-context learning ability of LLMs to solve unseen audio tasks, including audio understanding and generation tasks. Extensive experiments and ablation studies further validate the effectiveness of our method.
\end{itemize}


\section{Related works}
\textbf{Audio Codec Models} Historical investigations into low-bitrate parametric audio codecs began with earlier studies \citep{juang1982multiple,atal1971speech}; however, the quality of these codecs typically faced limitations. More recently, advancements have been made with the introduction of neural network-based audio codecs, evidenced by several promising developments \cite{soundstream,encodec,hifi-codec,dac,ns3}. These systems generally involve an encoder that extracts deep features within a latent space, which are then quantized and transmitted to a decoder. Particularly relevant to our work are the FACodec \cite{ns3} and SpeechTokenizer \cite{zhang2023speechtokenizer} models, which explicitly model different properties of audio in different vector quantization layers. Different from them, our proposed LLM-Codec tries to encode the audio data into a lexical word sequence.

\textbf{Multimodal Large Language Models}
Recently, there has been tremendous progress in the area of multimodal LLMs. These models use the pre-trained LLMs as the base model and try to take various modalities as additional input, such as vision \cite{minigpt5,liu2024language,yu2024spae,zhu2024beyond,li2023blip,tsimpoukelli2021multimodal}, audio \cite{chen2024salm,audio-flamingo,hu2024wavllm,zhang2023speechgpt,tang2023salmonn,hussain2023m,blsp}. In general, these multi-modal LLMs consist of a pre-trained LLM, a pre-trained vision/audio encoder, and a modality adaptor. They will construct a lot of multimodal datasets and use them to fine-tune the models. In the audio modality, most of the previous works try to solve speech understanding \cite{hu2024wavllm} or general audio understanding \cite{tang2023salmonn,hussain2023m}, and these models cannot apply to audio generation tasks. SpeechGPT supports a few audio understanding and generation tasks by fine-tuning all parameters and expanding the speech token's vocabulary based on LLAMA. However, the speech tokens in SpeechGPT only include semantic-level information, which limits its applications to more audio understanding and generation tasks. Furthermore, SpeechGPT does not explore the in-context learning ability to solve unseen tasks.

\textbf{In-context Learning}
In-context learning represents a form of few-shot learning, where a large language model (LLM) quickly adjusts to a specific task during inference by reviewing only a handful of examples provided in the prompt \cite{brown2020language}.  It has widely shown success in natural language tasks \cite{flan} and visual-language tasks \cite{alayrac2022flamingo,yu2024spae,zhu2024beyond,liu2024language}. In the audio domain, advanced methods have been proposed that utilize in-context learning to solve unseen audio tasks. SALM \cite{chen2024salm} proposes speech-augmented language models with in-context learning to solve speech recognition and speech translation tasks, they demonstrated that the SALM model can solve keyword boosting tasks. ICL-GSLM \cite{hsu2023exploration} proposes to use warmup training and prompt tuning strategies to empower the pre-trained speech language models \cite{gslm} in-context learning ability for unseen tasks. However, ICL-GSLM mainly focuses on exploring the in-context learning of audio understanding tasks and ignores audio generation tasks. 
Dynamic-superb \cite{huang2024dynamic} proposes to use instruction-tuning for audio understanding tasks. Similarly, \cite{wang2024bayesian} and \cite{wang2024can} also explore the in-context learning in the speech understanding domain. Inspired by the success in NLP tasks \cite{flan} and vision-language tasks \cite{yu2024spae,zhu2024beyond,liu2024language}, in this study, we focus on using the in-context ability from frozen LLMs to solve wide audio understanding and generation tasks.
\section{LLM-Codec}
\subsection{Overview}
\begin{figure*}[!t]
    \centering
    \includegraphics[width=\textwidth]{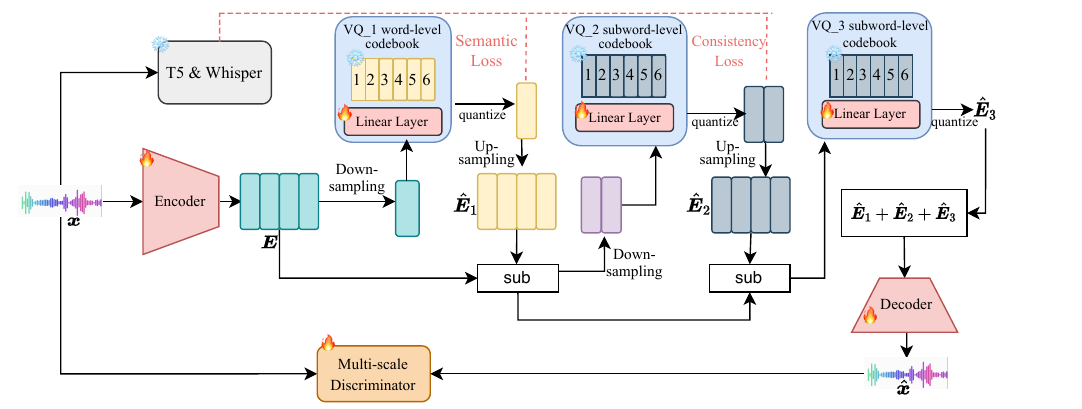}
    \caption{A high-level overview of LLM-Codec. Sub denotes the feature subtraction. We assume 3 RVQ layers are used in our study. In practice, we can use different RVQ layer settings.} 
    \label{fig:codec}
\end{figure*}

Previous audio codec models \cite{encodec,soundstream,hifi-codec} adopt a VQ-VAE \cite{van2017neural} framework to encode the audio signal into a discrete latent space, then decode the discrete token sequence into audio. Due to audio codec models mapping audio signal into a discrete token sequence, many works \cite{make-a-voice,uniaudio,borsos2023audiolm} propose to train an auto-regressive (AR) based language model to generate an audio token sequence by following the success of LLMs in natural language processing. But the discrete audio tokens produced by the codec model and text tokens in LLMs exist modal heterogeneity, \textit{e.g.} the codebooks in audio codec and the vocabulary of LLMs without any connection, which increases the difficulty of expanding well-trained LLMs to audio modality. Although previous works \cite{zhang2023speechgpt,rubenstein2023audiopalm} have demonstrated the effectiveness of expanding the vocabulary of LLMs to audio tokens and updating all of the parameters of LLMs, it will cost a lot of computing resources and forget the knowledge of the text. \\
In this part, we present a large language models-driven audio codec model (LLM-Codec), which effectively reduces the modal heterogeneity between audio and text. LLM-Codec is also based on the VQ-VAE framework, compared with previous work, the difference includes: (1) LLM-Codec is forced to quantize the audio signal into the token space of LLMs; (2) LLM-Codec adopts a multi-scale residual vector quantization strategy to balance the completeness and compactness of codec model; (3) LLM-Codec explicitly encodes different level information in different VQ layers. In the following, we give the details of LLM-Codec. Figure \ref{fig:codec} provides a visual depiction of the LLM-Codec.
\subsection{Encoder and Decoder} \label{codec-en-de}
For any audio $\boldsymbol{x}$, the encoder first encodes it into latent presentations $\boldsymbol{E}^{T*d}$, where $T$ denotes the number of frames, $d$ denotes the dimension of each vector. We set 4 down-sampling layers with $S=[3, 4, 5, 8]$ in the encoder, which results in 480 times down-sampling for audio. Then, each frame $\boldsymbol{e} \in \boldsymbol{E}$ is passed through the quantizer, which assigns it to the closest entry in a codebook, resulting in the quantized embedding $\boldsymbol{\hat{e}}$. Finally, the quantized feature $\boldsymbol{\hat{E}}$ inputs into the decoder to reconstruct $\boldsymbol{\hat{x}}$. Refer to Appendix \ref{appendix:codec} to find more model structure details. 

\subsection{Multi-scale residual vector quantization with the vocabulary of frozen LLM} \label{scrvq}
We use three residual VQ layers to maintain the balance between completeness and compactness. Furthermore, we propose to set different quantization granularity in different VQ layers: we expect the first VQ layer can encode the semantic information, and such information can be saved with fewer tokens, thus an interpolation function is used to down-sample the encoder features $\boldsymbol{E}^{T,d}$ into $\boldsymbol{E}_1^{T/k_1,d}$, then $\boldsymbol{E}_1^{T/k_1,d}$ is passed through the first VQ layer to obtain $\hat{\boldsymbol{E}}_1^{T/k_1,d}$. For the second VQ layer, we expect it can encode coarse-grained acoustic information, thus we pass the residual of the first VQ layer into the next VQ layer. Before that, we first up-sampling $\hat{\boldsymbol{E}}_1^{T/k_1,d}$ into $\hat{\boldsymbol{E}}_1^{T,d}$, then obtain the residual features by 
\begin{equation}
 \boldsymbol{E}_2^{T,d} = \boldsymbol{E}^{T,d} - \hat{\boldsymbol{E}}_1^{T,d}.
\end{equation}
Similarly, we also apply a down-sampling operation to $ \boldsymbol{E}_2^{T,d}$, we set the down-sampling step as $k_2$.  The features become as $\boldsymbol{E}_2^{T/k_2,d}$. Then we pass it into the second VQ layer and obtain $\hat{\boldsymbol{E}}_2^{T/k_2,d}$. Lastly, we expect the last VQ layer can preserve all of the residual acoustic information. We first obtain the residual features based on the quantized features of the first two VQ layers
\begin{equation}
 \boldsymbol{E}_3^{T,d} = \boldsymbol{E}^{T,d} - \hat{\boldsymbol{E}}_1^{T,d} - \hat{\boldsymbol{E}}_2^{T,d}.
\end{equation}
Considering the residual acoustic information is more complex and diverse, we directly apply the VQ operation to each frame without any down-sampling. By using a large down-sampling step in the encoder of codec, and applying a multi-scale VQ strategy, we can effectively reduce the number of quantized audio token sequences. In our setting, 1-second audio with a 16k sampling rate will be quantized into 57 tokens. To ensure that the first VQ layers encode semantic information, we propose incorporating a semantic loss during the training process. Furthermore, to maintain the training stability, we propose a consistency loss. The details will be introduced in Section \ref{train loss}. \\
\textbf{The initialization of VQ layers}
To generate lexical tokens, we utilize a pre-trained LLAMA 2 codebook to initialize the VQ layers. Considering that the first layer, the VQ layer, is designed to encode the semantic information, we do not directly use the full LLAMA codebook. Instead, we define a new codebook based on Oxford 5000 Words, these words are commonly used to make up any meaningful sentence. We choose these words that only consist of one or two sub-words in the LLAMA codebook. If a word includes two sub-words, we use the mean representation of two sub-words in the LLAMA codebook as the final representation. Lastly, the codebook size of the first VQ layer is 3248. We directly use the LLAMA codebook to initialize the second and third VQ layers. The codebook size is 32000. Furthermore, the LLAMA codebook embedding dimension is 4096, which is too large for codec training. Thus, we apply a linear mapping to 512. In the training process, the parameters of codebooks are fixed.

\subsection{Training loss} \label{train loss}
Our approach is based on a GAN objective, in which we optimize both the generator(it consists of encoder, quantizer, and decoder) and the discriminators. For the generator, its training loss consists of three parts: (1) reconstruction loss term; (2) adversarial loss term (via discriminators); and (3) semantic and consistency losses. In the following, we give the details of proposed semantic loss and consistency loss. Refer to Appendix \ref{appendix:rec_loss} to find the details of reconstruction loss and adversarial loss.

\textbf{Semantic loss}
To enhance the semantic representation ability in the first layer, we introduce a semantic loss for the first VQ layer. We expect it can encode semantic information, for example, if the input audio includes a sound event, the first layer should encode which semantic information of the sound event. Similarly, if the input audio is speech, the first layer should encode the content of the speech. To realize this target, we use a pre-trained T5-base model \cite{t5} to extract a global representation vector $\boldsymbol{g}$ for the input audio content. We use Whisper to obtain its transcriptions if the input audio is speech. If the input audio is sound, we use its audio caption label:
\begin{equation}
\mathcal{L}_{s} = L_1(mean(\hat{\boldsymbol{E}}_1^{T,d}), \boldsymbol{g} )
\end{equation}
\textbf{Consistency loss} In our early experiments, we found the training of LLM-Codec is not stable, and the model is easy to collapse. One of the reasons is that we designed a significant down-sampling rate and the codebooks are fixed in the training, which increases the training difficulty. To solve this issue, we propose a consistency loss to maintain the training stability. Specifically, we propose using a pre-trained Whisper encoder \cite{whisper} to extract frame-level features $\boldsymbol{w}$, then using these features as prior knowledge to guide the second VQ layer. 
\begin{equation}
\mathcal{L}_{c} = L_1(\hat{\boldsymbol{E}}_2^{T/2,d}, inp (\boldsymbol{w}))
\end{equation}
where $inp$ denotes the interpolation function to align the feature dimension between the quantized features and whisper features. We chose the Whisper encoder because it is trained not only on speech data but also on non-speech data. Furthermore, we do not apply this loss on the third VQ layer, because we expect the third VQ layer to encode the residual information. 

\begin{table*}[t]
    \centering
    \small
    \caption{Performance comparison between open-sourced audio codec models, baselines, and the proposed LLM-Codec. Evaluation is conducted on the VCTK dataset \cite{Veaux2017CSTRVC}.}
    \vspace{2mm}
    \begin{tabular}{lcc|cc}
    \toprule
    Model   & Down-sampling steps & Tokens per second  & PESQ  & STOI  \\
    \midrule
    Encodec\_24k (3 Vanilla RVQ) \cite{encodec}   & 320 & 225 & 2.18  & 0.79        \\
    DAC\_16k (3 Vanilla RVQ) \cite{dac}    & 320 & 150 & 1.76 & 0.78     \\
    \midrule
    Baseline (3 Vanilla RVQ)    & 480  & 99  & 2.64 & 0.83   \\
    Baseline (2 Multi-scale RVQ)   & 480 & 41 & 2.22 & 0.79   \\
    Baseline (1 VQ)   & 480 & 33 & 2.01 & 0.76      \\
    LLM-Codec (Ours)    & 480  & 57  & 2.55 & 0.82   \\
    \bottomrule 
    \end{tabular}
    \label{tab:codec}
\end{table*}
\begin{table}[tp]
\centering
\caption{Audio understanding task evaluation results. Task induction denotes the explanatory text that precedes the sequence of audio and text. It is intended to describe the task to the model in natural language, for example: Please answer the question. Accuracy (\%) is used as the metric. For the Random guess, we calculate the average based 5 times evaluation. K shots refers to the number of distinct samples for each category, and Repeats refer to how many times we copy the prompt samples.}
\label{tab:speech_emotion}
\begin{tabular}{lllcccccc}
\toprule
\multicolumn{3}{r}{Task Induction} &  \usym{2613}   & \checkmark & \checkmark& \checkmark& \checkmark \\
Method &\# Layers  & \multicolumn{1}{r}{K Shots}   & 1    & 1    & 3       & 1    & 1        &                      \\
&  & \multicolumn{1}{r}{Repeats}         & 0    & 0    & 0       & 2    & 3                           \\
\midrule
\multicolumn{7}{l}{\textit{2-way speech emotion classification}} \\
Random  & None & & \multicolumn{3}{r}{44}  &  &  &             \\
BLSP \cite{blsp} & Whisper encoder &  & 9 & 29 & 50 & 33 & 19          \\
UniAudio 1.5 (ours) & semantic layer &  & 25 & \textbf{53} & \textbf{59} & 53 & 54           \\
UniAudio 1.5 (ours) & semantic + acoustic layers &  & \textbf{45} & 49 & 53 & \textbf{55} & \textbf{54}            \\
\midrule
\multicolumn{7}{l}{\textit{2-way sound event classification.}} \\
Random  & None & & \multicolumn{3}{r}{45}  &  &  &             \\
BLSP \cite{blsp} & Whisper encoder &  & 44 & 47 & 54 & 15 & 17    \\
UniAudio 1.5 (ours) & semantic layer &  & \textbf{48} & \textbf{60} & \textbf{57} & \textbf{57} & \textbf{73}                 \\
UniAudio 1.5 (ours) & semantic+acoustic layers &  & 41 & 48 & 55 & 54 & 62                \\
\midrule
\multicolumn{7}{l}{\textit{3-way sound event classification.}} \\
Random  & None & & \multicolumn{3}{r}{30}  &  &  &             \\
BLSP \cite{blsp} & Whisper encoder &  & 23 & 26 & 36 & 24 & 16            \\
UniAudio 1.5 (ours) & semantic layer &  & \textbf{38} & \textbf{41} & \textbf{39} & 43 & 42                \\
UniAudio 1.5 (ours) & semantic+acoustic layers &  & 25 & 37 & 35 & \textbf{44} & \textbf{50}                 \\
\bottomrule
\end{tabular}
\end{table}

\section{UniAudio 1.5}
By combining the pre-trained LLMs and the proposed LLM-Codec models, we can solve many audio tasks in a few-shot style, as Figure \ref{fig:undersanding} shows. We named the system as UniAudio 1.5 for the reason that the system can be viewed as a universal audio task solver. 
\subsection{Connection to UniAudio}
UniAudio 1.5 is an advanced edition of the UniAudio Series \cite{uniaudio}. Compared to its previous
version UniAudio \cite{uniaudio}, UniAudio 1.5 has the following connections and distinctions. First,
\textbf{goal}. While both UniAudio 1 and UniAudio 1.5 aim at building a universal audio foundation model for all audio tasks, their focuses are different. UniAudio focuses on audio generation tasks, such as text-to-speech, text-to-music, singing voice generation, and so on. UniAudio 1.5 focuses on audio understanding and generation tasks by exploring the few-shot
ability based on large language models. Second, \textbf{architecture}.
UniAudio 1.5 keeps the basic components in UniAudio, such as an audio codec used to transfer the audio modality into discrete representations, and a decoder-only transformer backbone is used.
However, UniAudio 1.5 leverages 1) a pre-trained LLMs to solve the audio understanding and generation tasks by in-context learning, 2) a LLM-driven audio codec to quantize
the audio data into the token space of LLMs.  

Building a multi-modal audio foundation model that is capable of handling any audio task is the ultimate goal of the UniAudio series. In UniAudio 1.0, we show the possibility to build a universal model for different types of audio generation tasks, but it (1) cannot effectively solve audio understanding tasks; (2) cannot solve unseen audio tasks in the training or fine-tuning stages. UniAudio 1.5 shows the possibility to use pre-trained LLMs for both audio understanding and generation tasks. We believe the proposed LLM-Codec in UniAudio 1.5 builds a foundation for more advanced editions of the UniAudio Series in the future.

\section{Experimental Results}

\subsection{Experimental Settings}
\textbf{Training data} LLM-Codec is a universal audio codec model, we train it on both speech and sound datasets. For speech data, we use part of the MLS dataset \cite{mls}. For sound data, we use the AudioCaps dataset \cite{kim2019audiocaps}. In total, we use 2k hours of audio data to train the LLM-Codec model. \\
\textbf{Model setting} As described in \ref{codec-en-de}, the encoder and decoder of LLM-Codec consist of several Convolution and Transformer blocks. For the quantizer, we use three residual vector quantization layers. The down-sampling rate for the first two layers is set as $k_1=4$ and $k_2=2$. We initialize the parameters of VQ layers with the help of the LLAMA2 7B model's vocabulary. Considering the latent dimension of LLAMA2 is 4096, we use a learnable linear layer to map them into 512. \\
\textbf{Evaluation metrics} To verify the reconstruction performance of the LLM-Codec, Perceptual Evaluation of Speech Quality (PESQ) and Short-Time Objective Intelligibility (STOI) are used. For audio understanding tasks, we conduct a lot of N-way-K shot classification experiments, and use accuracy as the metric. For the audio generation task, we follow commonly used metrics in each task. \\
\textbf{Evaluation dataset} We choose the commonly used test dataset for each task and construct N-way-K-shot test pairs. More details about construct evaluation samples can be found in Appendix \ref{appenxid-n-way}.\\ 
\textbf{Baselines} Given the limited number of works that focus on exploring few-shot learning for unseen audio tasks, we choose BLSP \cite{blsp} as one of the baselines for audio understanding tasks.  Since BLSP is fine-tuning with a continuation writing task and does not explicitly introduce audio classification tasks, thus these audio classification tasks are unseen for the BLSP model. Furthermore, we also compared with the instruction-tuning-based models in dynamic-superb \cite{huang2024dynamic}. For audio generation tasks, we do not find related works, thus we report the performance of state-of-the-art special models. 
\subsection{Main results}
We first present the reconstruction performance comparison. Then we apply the LLM-Codec and LLAMA 2 7B model (named as UniAudio 1.5) for audio understanding and audio generation tasks, to verify the ability of the proposed method. Lastly, we give the visualization of LLM-Codec to explain why it can work. We leave more experiments on Appendix \ref{appendix:more}. \\
\textbf{Reconstruction performance}
We compare the audio reconstruction quality with previous works Encodec \cite{encodec}, DAC-Codec \cite{dac}, and our baseline model. We report Perceptual Evaluation of Speech Quality (PESQ) and Short-Time Objective Intelligibility (STOI). Table \ref{tab:codec} shows the results. Compared to previous methods, the LLM-Codec achieves better reconstruction performance while utilizing fewer tokens. More specifically, the LLM-Codec model can compress 1-second audio data into a sequence that only includes 57 tokens, which significantly reduces the sequence length. Compared to the RVQ baseline model, the LLM-Codec significantly reduces the compressed tokens, and its reconstruction performance does not significantly decline. In Section \ref{sec:ablation}, we will show the importance of compressing audio into fewer tokens. We also conduct experiments to validate whether we can use a few VQ layers, such as 1 VQ layer or 2 VQ layer, we can see that the reconstruction performance will significantly drop. To maintain the balance between completeness and compactness, we choose a multi-scale 3 VQ layer as the default setting.
\begin{table*}[t]
    \centering
    \small
    \caption{Evaluation on dynamic-superb benchmark tasks. Accuracy (\%) is used as the metric.}
    \vspace{2mm}
    \begin{tabular}{lcccc}
    \toprule
    Task   & ImageBind-LLM \cite{huang2024dynamic} & Whisper-LLM \cite{huang2024dynamic}   & ASR-ChatGPT \cite{huang2024dynamic}  & Ours  \\
    \midrule
    Accent Classification   & 19 & 4  & 7  & \textbf{24}        \\
    Bird Sound Detection    & 28 & 14  & 15 & \textbf{50}     \\
    Chord Classification    & 44  & \textbf{58}   & 3 & 55   \\
    Language Identification   & 26  &  13 & \textbf{96} &  25   \\
    \bottomrule 
    \end{tabular}
    \label{tab:dynamic}
    \vspace{-0.6 cm}
\end{table*}
\begin{wraptable}{r}{7cm}
    \centering
    \small
    \caption{Text-to-speech generation performance.}
    \vspace{2mm}
    \scalebox{1.0}{
    \begin{tabular}{l|cc}
    \toprule
    Model    & ACC  & DNSMOS  \\
    \midrule
    GT  & -  & 2.91        \\
    FastSpeech 2  & -  & 3.42        \\
    \midrule
    UniAudio 1.5 (Ours) & 70 & 2.92   \\
    \bottomrule 
    \end{tabular}
    \label{tab:tts} }
\end{wraptable}
\textbf{Speech Emotion Classification}
The speech emotion classification task \cite{speech-emotion-survey} aims to predict the emotion label of the speech. We conduct 2-way K-shot experiments on the ESD \cite{esd} dataset. Experimental results are shown in Table \ref{tab:speech_emotion}. We have the following findings: (1) Task induction is important to maintain the stability of performance, we can see that without task induction, the classification accuracy will dramatically decline. (2) The semantic layer effectively extracts the global semantics of audio, which can be easily understood by the LLAMA model. (3) Using more demonstration samples (\textit{e.g.} 3 shots), the performance will be better. (4) Repeating the demonstration samples can also bring improvement. (5) Compared to the BLSP, our method performs better in any setting. Furthermore, we also note that the performance of BLSP will drop when repeat operation is used. One possible reason is that BLSP only learns the translation relationship between text and speech, repeating samples cannot bring new cues for LLMs to solve the new task. Instead, our LLM-Codec learns to map the audio data into the latent space of LLMs, increasing the number of demonstration samples can help LLMs to find special patterns to solve this new task. 

\textbf{Sound Event Classification}
Sound event classification aims to recognize the sound event in the audio. In general, an audio may include multiple events. To simplify the recognition difficulty, we assume each audio only includes one event. We conduct experiments on the ESC50 dataset \cite{esc50}, which includes 50 different types of events. We construct 2-way-K-shot and 3-way-K-shot evaluations based on the ESC50 test set. Compared with the BLSP model, our proposed method gets better performance. Based on the experimental results from two audio understanding tasks, we can see that the semantic VQ layer is very important for understanding tasks. \\
\textbf{Dynamic-SUPERB Benchmark} We also conduct experiments on Dynamic-SUPERB Benchmark tasks \cite{huang2024dynamic}. In \cite{huang2024dynamic}, authors propose an instruction-tuning strategy for multi-modal LLMs. They first construct a lot of audio tasks as training data, then validate some unseen audio tasks in a zero-shot way. To make a fair comparison, we use the same test set with them, and choose the first N samples as the demonstration to construct a N-way-1-shot evaluation. As Table \ref{tab:dynamic} shows 4 selected audio understanding tasks, our proposed method obtains better or compared performance than these baselines in \cite{huang2024dynamic}. Especially, for the bird sound detection task, our proposed method obtained great improvement over previous methods. We also note that our method performs worse on language identification, the possible reason is that our codec model is only trained on English speech data. In the following, we will show that UniAudio 1.5 also can be used to conduct audio generation tasks. 

\begin{figure*}[t]
    \centering
    \includegraphics[width=\textwidth]{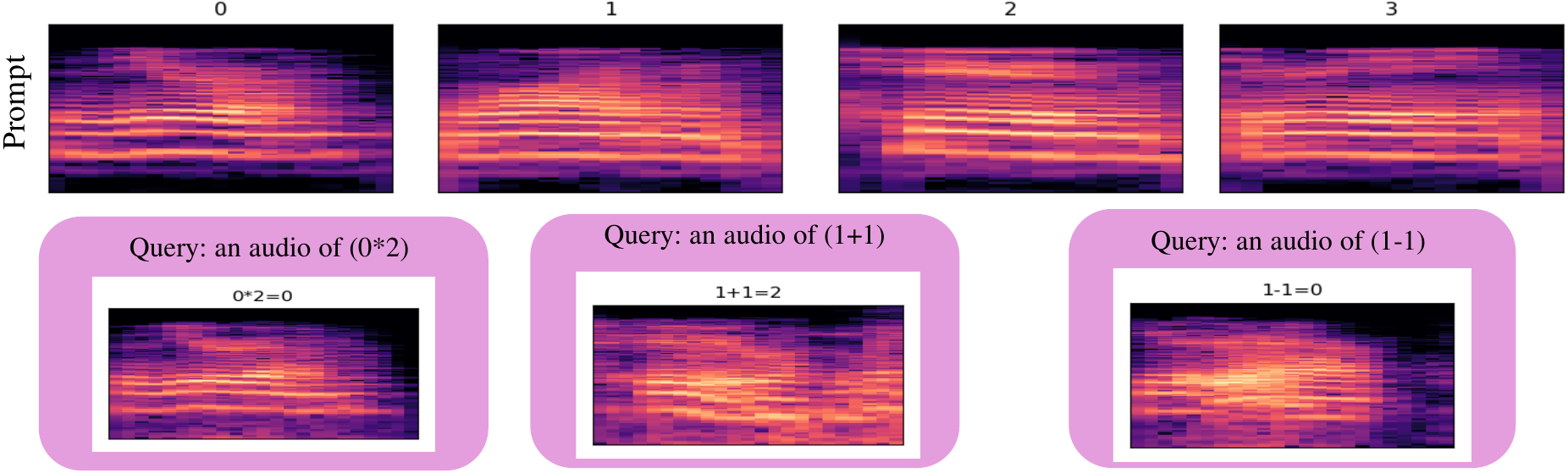}
    \caption{Examples of simple text-to-speech generation using LLM-Codec and LLAMA2 model.} 
    \label{fig:tts_ex}
\end{figure*}

\textbf{Simple text-to-speech generation}
We conduct text-to-speech generation on the Free Spoken Digit Dataset (FSDD) dataset \cite{speech_digit}, which includes 3 speakers and 1,500 recordings. Unlike the traditional TTS model, which generates any speech content, this task generates digit speech. Our current model to generate complex speech content is still challenging. We use accuracy (ACC) to assess the content of the generated sample whether following the instructions. DNSMOS is used to assess the speech quality of generated samples. We construct 20 different query questions, including addition, subtraction, multiplication, division, and reasoning (finding more details from Appendix \ref{appenxix:tts}). From Table \ref{tab:tts}, we can see that our proposed model can accurately understand the query in most cases (the accuracy is 70 \%) and generate good-quality speech samples. Figure \ref{fig:tts_ex} gives a visualization of generating speech based on the query. The frozen LLAMA model learns about 4 digits (0-3), each audio digit includes 5 samples. We add the context for each audio: "an audio of k" before inputting the audio's discrete representations into LLAMA, as Figure \ref{fig:undersanding} (b) shows. After that, we let the LLAMA 2 model generate corresponding speech digits based on the instruction. We also note that the generated audio appears different from all context audio samples, demonstrating the cross-modal reasoning capability of LLMs when using the LLM-Codec as the connector for text and audio.  

\begin{figure*}[t]
    \centering
    \includegraphics[width=\textwidth]{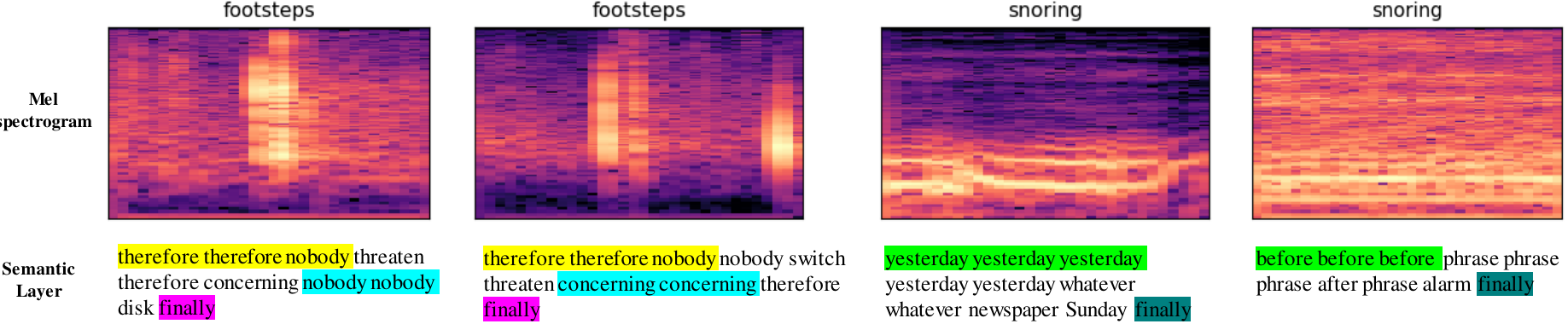}
    \caption{The token visualization with LLM-Codec. The audio samples are from the ESC50 dataset.} 
    \label{fig:vis}
\end{figure*}

\begin{wraptable}{r}{6cm}
    \centering
    \small
    \caption{Speech denosing evaluation.}
    \vspace{2mm}
    \scalebox{1.0}{
    \begin{tabular}{l|cc}
    \toprule
    Model    & PESQ  & STOI  \\
    \midrule
    SGMSE+ \cite{richter2023speech}  & 3.53  & 0.79        \\
    \midrule
    UniAudio 1.5 (Ours)   & 2.17 & 0.57   \\
    \bottomrule 
    \end{tabular}
    \label{tab:se} }
\end{wraptable}
\textbf{Simple Speech Denoising}
To verify whether the proposed method can conduct speech-denoising tasks, we simulate noisy speech based on the VCTK dataset and NoiseX-92 dataset, we set the SNR ranges from -20 to 20. For each clean speech, we choose 5 different noises to simulate noisy speech. The first 4 noisy and clean audio pairs are used as demonstrations, and the model learns to denoise the last noisy one. To improve in-context learning ability, we repeat the demonstration samples 4 times. The experimental results as Table \ref{tab:se} shows, we can see that the proposed method can also learn to denoise without any training. Furthermore, we also note that the performance has a large room to improve compared to special models. \\
\textbf{Token Visualization} We visualize the tokens produced by the first VQ layer of LLM-Codec for different types of sound in Figure \ref{fig:vis}. We have the following findings: (1) Although the two audios include the same sound event, the quantized sequence is not exactly the same. (2) The quantized sequence of two same types of audio has a similar pattern, \textit{e.g.} their token sequences have similar repeating patterns or the same word. Such patterns may help the LLMs recognize the type of audio.

\subsection{Ablation study} \label{sec:ablation}
\textbf{The influence of multi-scale RVQ} We first conduct experiments to see the effectiveness of multi-scale RVQ. As Table \ref{tab:ablation} shows, compared with vanilla RVQ, the proposed multi-scale RVQ does not bring a significant reconstruction performance drop, which validates our assumption that semantic information does not need too much token to encode. Secondly, we find the multi-scale RVQ significantly reduces the length of the token sequence and brings benefits for downstream tasks (the audio classification accuracy is better than the baseline). One potential explanation is that LLMs can better identify the unique pattern in a brief sequence. Intuitively, the semantic information included in a 1-second audio is limited. It is unnecessary to use very long sequences to represent limited information. \\
\textbf{The influence of down-sampling times} We can see that using a smaller down-sampling rate (320) can improve the reconstruction performance, but it also increases the length of the token sequence. We can see that the classification accuracy will decrease when the sequence length increases. \\
\textbf{The influence of semantic loss}
Without semantic loss, the performance of the audio understanding task will drop. Furthermore, we also find that adding semantic loss does not influence the reconstruction performance. In summary, the proposed semantic loss is very useful. \\
\textbf{The influence of consistency loss} We find that consistency loss is important to maintain training stability. Without it, we can see the model fails to reconstruct the audio. We conjecture that frozen codebooks and large compression rates significantly improve the difficulty of training. The consistency loss forces the second VQ layer to produce features similar to those of the Whisper encoder, which provides guidance for vector quantization and prevents the model from collapsing in the early stage.  \\
\textbf{The influence of word-level codebooks} We also conduct experiments to show the effectiveness of using word-level codebooks to initialize the first VQ layer. Compared with using sub-word vocabulary for the first VQ layer, we can see that using the proposed word-level codebook can improve the reconstruction performance and classification accuracy.  \\
\textbf{The importance of frozen codebooks} LLM-Codec compresses the audio data into the token space of LLMs by initializing the codebooks with the LLMs' vocabulary and fixing it during the training stage. Table \ref{tab:ablation} also presents the results of updating codebooks: it can improve the reconstruction performance, but the accuracy is a significant drop. The result is consistent with our hypothesis: updating the codebooks parameter will decrease the codec training difficulty, but the learned codebook space is different from the LLM's token space, resulting in the downstream task performance declines. \\
\textbf{Different setting of $k_1$ and $k_2$ in multi-scale RVQ} We validate a new setting for multi-scale RVQ with $k_1=3$ and $k_2=5$. We can see that the reconstruction performance will decline. We think one of the reasons is that the second VQ layer should not apply a large down-sampling step, which significantly influences the reconstruction.  \\
\textbf{Codebook usage} Previous works \cite{huh2023improvedvqste,soundstream,dac} suggest that using a large-scale codebook may result in codebook collapse ((where a fraction of the codes are unused). We calculate the codebook usage for each VQ layer in LLM-Codec. The used codes are 3246 (3248), 31911 (32000), 31941 (32000) for each VQ layer, which shows that most of codes are used.

\begin{table*}[h]
    \centering
    \small
    \caption{Ablation studies on training loss, multi-scale RVQ setting, initialization of VQ layer. The classification accuracy (\%) is evaluated under the sound event classification task 2-way 1-shot setup.}
    \vspace{2mm}
    \begin{tabular}{lcc|ccc}
    \toprule
    Model   & Down-sampling & Tokens per second  & PESQ  & STOI & ACC \\
    \midrule
    Baseline (Vanilla 3 RVQ)   & 480 & 99 & 2.64  & 0.83 & 55    \\
    LLM-Codec (Multi-scale 3 RVQ)   & 480 & 57 & 2.55 & 0.82 & 60     \\
    LLM-Codec (Multi-scale 3 RVQ)   & 320 & 87 & 2.60 & 0.83 & 57     \\
    \midrule
    w/o semantic loss & 480 & 57 & 2.54 & 0.82 & 58     \\
    w/o consistency loss & 480 & 57 & 1.19 & 0.53 & 48     \\
    w/o word-level codebook & 480 & 57 & 2.46 & 0.81 & 59     \\
    updating codebooks & 480 & 57 & 2.63 & 0.83 & 55     \\
    seting $k_1=3$ and $k_2=5$ & 480 & 50 & 2.35 & 0.79 & 58    \\
    \bottomrule 
    \end{tabular}
    \label{tab:ablation}
\end{table*}

\section{Conclusion}
In this study, we explore a cross-modal in-context learning approach to solve unseen audio tasks in a few-shot style. Specifically, we propose to train a LLMs-driven audio codec (LLM-Codec) that compresses the audio signal into the token space of LLMs. The LLM-Codec effectively reduces the modal heterogeneity between text and audio. With the help of LLM-Codec, pre-trained LLMs can be applied to audio understanding and generation tasks. We demonstrate that LLM-Codec has good reconstruction performance, and the compressed token sequence is suitable for LLMs to understand and generate. Experiments show that the LLMs equipped with the proposed LLM-Codec, prompted by only a few examples, are capable of achieving the expected functions in many scenarios.

\section{Limitations}
Although we show the possibility of using the in-context learning ability of LLMs for unseen audio tasks without any parameter update, the performance of these tasks is still poorer than these special models in the audio domain. The capability to learn within an in-context framework is significantly limited for a modality that was not exposed during the training process. Due to the LLM's context length limitation, we cannot add more demonstration samples to help improve the performance. We think it is worth exploring using more demonstrations to improve its in-context learning ability. Moreover, we only explore to use LLAMA 2 7B model as the backbone, more advanced open-sourced LLMs are worth exploring. Considering we have open-sourced the LLM-Codec, readers can conduct experiments on their favorite LLMs. In the future, we will explore to train multi-modal LLMs by fine-tuning LLMs on text-audio datasets with the help of LLM-Codec.

\bibliographystyle{plain}
\bibliography{neurips_2023}
\newpage
\appendix
\begin{center}{\bf {\LARGE Appendices} }
\end{center}
\section{Appendix Overview}
These Appendices provide additional details to support our main manuscript, including (1) the training detail and model structure of LLM-Codec. (2) The details of the evaluation dataset. (3) More audio task evaluation results. (4) Limitations.
\section{More details of LLM-Codec} \label{appendix:codec}
\subsection{Model structure}
Table \ref{tab:codec-config} gives the details of LLM-Codec configuration, which results in 160M parameters. To facilitate research on cross-modal in-context learning and multi-modal LLMs, we have open-sourced the LLM-Codec models.
\begin{table}[htbp]
  \centering
    \begin{tabular}{c|c}
    \toprule
           & LLM-Codec \\
    \midrule 
    Input shape & (1, 1, T) \\
    Encoder (input dimension) & 32 \\
    Down-sampling rate & [3, 4, 5, 8] \\
    latent dimension  & 512 \\
    Codebook dimension & 4096 \\
    Transformer layer dimension & 512 \\ 
    Number of Transformer heads & 8 \\
    Decoder dimension  & 1536 \\
    Up-sampling rate  & [8, 5, 4, 3] \\
    VQ strides  & [5, 3, 1] \\
    \bottomrule
    \end{tabular}%
  \vspace{5pt}
  \caption{LLM-Codec model backbone configurations}
  \label{tab:codec-config}%
\end{table}

\textbf{Encoder and Decoder}
 Considering a single-channel audio signal $\boldsymbol{x} \in \mathcal{R}^{t \times sr}$, where $t$ and $sr$ denote the audio duration and the sample rate. The overall architecture is similar to previous audio codec models, such as Encodec \cite{encodec}, DAC \cite{dac}, and HiFi-Codec \cite{hifi-codec}, which includes four main parts: encoder, quantizer, decoder, and discriminators. Figure \ref{fig:codec} provides a visual depiction of the proposed method. For any input $\boldsymbol{x}$, the encoder first encodes it into latent presentations $\boldsymbol{E}^{T*d}$, where $T$ denotes the number of frames, $d$ denotes the dimension of each vector. Due to the encoder includes some down-sampling layers, resulting in $T << t \times sr$. Then each frame $\boldsymbol{e} \in \boldsymbol{E}$ is passed through the quantizer, which assigns it to the closest entry in a codebook, resulting in the quantized embedding $\boldsymbol{\hat{e}}$. Finally, the quantized feature $\boldsymbol{\hat{E}}$ inputs into the decoder to reconstruct $\boldsymbol{\hat{x}}$. The encoder and decoder architecture follows previous works Encodec \cite{encodec} and DAC-Codec \cite{dac}, which includes several convolution layers and transformer layers. Specifically, the encoder model comprises a 1D convolution with $C$ channels and a kernel size of 7, leading into $B$ convolution blocks. Each block contains a residual unit followed by a down-sampling layer, which employs a convolution with a kernel size $K$ that is twice the stride $S$. The residual unit itself comprises two convolutions, each with a kernel size of 3, linked by a skip connection. The transformer block is used for sequence modeling, and concludes with a final 1D convolution layer featuring a kernel size of 7. In this study, we set $S=[3, 4, 5, 8]$, which results in 480 times down-sampling for audio. The decoder mirrors the encoder's architecture, substituting stride convolutions with transposed convolutions and reversing the stride order.

\textbf{Discriminators} For the discriminators, we follow previous work \cite{uniaudio}, which combines the mel-spectrogram and log-mel-spectrogram features and then input them into a network consisting of several convolutional layers. In our experiments, we use 6 different discriminators with different configurations. Specifically, we set the hidden dimension as \{64, 128, 256, 512, 512, 512\} and the hop length as \{32, 64, 128, 256, 512, 1024\}. \\
\subsection{Reconstruction loss and adversarial loss for LLM-Codec} \label{appendix:rec_loss}
The reconstruction loss is calculated between $\boldsymbol{x}$ and $\hat{\boldsymbol{x}}$. We design the loss from two aspects: the time domain and the frequency domain. For the time domain, we directly calculate the $L_1$ loss between $\boldsymbol{x}$ and $\hat{\boldsymbol{x}}$. For the frequency domain, we calculate the $L_1$ loss between the STFT spectrogram of $\boldsymbol{x}$ and $\hat{\boldsymbol{x}}$. Note that a sub-band split strategy \cite{wang2024consistent} is used to split the spectrogram into several parts, and then we calculate the loss between these sub-bands. The adversarial loss is used to improve the perceptual quality of generated audio. A multi-scale Mel-spectrogram discriminators \cite{uniaudio} is used. To train the discriminator, we can optimize the following objective function:
\begin{align}\label{dis loss}
    \mathcal{L}_{\mathit{d}} = \frac{1}{K} \sum_{i=1}^K \mathit{max}(0, 1-D_k(\boldsymbol{x}))+\mathit{max}(0,1+D_k(\boldsymbol{\hat{x}})) 
\end{align}
where $K$ denotes the number of discriminators. In the training stage, the adversarial loss for the generator is calculated as a hinge loss over the logits of these discriminators:
\begin{align}\label{adv loss}
    \mathcal{L}_{\mathit{adv}} = \frac{1}{K} \sum_{i=1}^K \mathit{max}(0, 1-D_k(\boldsymbol{\hat{x}}))
\end{align}
We also compute the feature loss by taking the average absolute difference between the discriminator's internal layer outputs for the generated audio and those for the corresponding real audio.
\subsection{Training details}
The AdamW optimizer is used in the training. We set the learn rate as $1e-4$. We train the model with 100k steps. For the training loss, we combine all of the loss terms without a special loss design. In the training stage, we use the pre-trained T5-base model and Whisper-base model for the reason that their latent dimension is both 512. We conduct all of the experiments with 2 NVIDIA A100-80G GPUs.
\section{Evaluation dataset}
In this part, we show how to construct an evaluation dataset for the N-way-k-shot test. 
\subsection{N-way-k-shot test samples} \label{appenxid-n-way}
\paragraph{Speech emotion classification with LLAMA 2.}
We give an example of 2-way 1-shot classification tasks. Firstly, we get the emotion class set from the ESD dataset: ['Angry', 'Happy',  'Neutral',  'Sad', 'Surprise']. Then we randomly choose two emotions as targets, and get the corresponding audios. For example, assuming that we get \emph{Happy} and \emph{Sad}
the prompt can be
\begin{Verbatim}
For each of the following input-output pairs, the output is 
one of [`Happy' or `Sad']
###
Input: <token sequence from a happy emotion of audio>
Output: happy
###
Input: <token sequence from a sad emotion of audio>
Output: sad
###
Input: <token sequence from the query audio>
Output:
\end{Verbatim}
We use greedy decoding to get a maximum of 16 tokens from LLAMA 2 7B.
\paragraph{Sound event classification with LLAMA 2.}
We give an example of 3-way 1-shot classification tasks. Firstly, we get the sound event class set from the ESC50 dataset. Then we randomly choose three sound events as targets, and get the corresponding audio. For example, assuming that we get \emph{dog}, \emph{speaking}, and \emph{mouse click}
we set the prompt as
\begin{Verbatim}
For each of the following input output pairs, 
output is one of [`dog' or `speaking' or 'mouse_click']
###
Input: <token sequence from a dog event of audio>
Output: dog
###
Input: <token sequence from a speaking event of audio>
Output: speaking
###
Input: <token sequence from a mouse click event of audio>
Output: mouse click
###
Input: <token sequence from the query audio>
Output:
\end{Verbatim}
\subsection{Audio generation} \label{appenxix:tts}
\paragraph{Text-to-speech generation with LLAMA 2}
\begin{Verbatim}
Instruction: Learn a foreign language for different digits, 
then generate the corresponding number using 
foreign language based on instruction
###
Input: <an audio of 1>
Output: <token sequence of audio 1>
###
Input: <an audio of 2>
Output: <token sequence of audio 2>
###
Input: <an audio of 3>
Output: <token sequence of audio 3>
###
Input: <an audio of 1+1>
Output:
\end{Verbatim}
To simplify to generation process, we set each audio has the same duration. 
\paragraph{Text-to-speech question design}
we designed 20 different questions for text-to-speech, which include addition, subtraction, multiplication, division, and reasoning. 
\begin{Verbatim}
###
Input: <an audio of (1+1)>
###
Input: <an audio of (1+2)>
###
Input: <an audio of (2+2)>
###
Input: <an audio of (5-1)>
###
Input: <an audio of (5-2)>
###
Input: <an audio of (1-1)>
###
Input: <an audio of (0*2)>
###
Input: <an audio of (2*2)>
###
Input: <an audio of (1/1)>
###
Input: <an audio of (2/1)>
###
Input: <an audio of (4/2)>
###
Input: <an audio of (the square root of 4)>
###
Input: <an audio of (the square root of 1)>
###
Input: <an audio of (the last digit of 110)>
###
Input: <an audio of (the first digit of 110)>
###
Input: <an audio of (the sum of 1+1+1)>
###
Input: <an audio of (the next digit of 4)>
###
Input: <an audio of (sequence 0,1,2,3 what is next?)>
###
Input: <an audio of (sequence 4,3,2,1 what is next?)>
###
Input: <an audio of (how many days in a week)>

\end{Verbatim}
\section{More audio tasks evaluation experiments with the proposed method} \label{appendix:more}
In the following, we show the results of speech command recognition and text-to-sound generation. 
\subsection{Speech Command Recognition}
\begin{table}[h]
\centering
\caption{Speech command recognition evaluation results on Speech Command dataset. Accuracy (\%) is used as the metric. For the Random guess, we run 5 times then calculate the average.}
\label{tab:speech_command}
\begin{tabular}{lllcccccc}
\toprule
\multicolumn{3}{r}{Task Induction} & \usym{2613}   & \checkmark & \checkmark& \checkmark& \checkmark  \\
Method &\# Layers & \multicolumn{1}{r}{K Shots}   & 1    & 1    & 3       & 1    & 1                              \\
&  & \multicolumn{1}{r}{Repeats}         & 0    & 0    & 0       & 1    & 3                         \\
\midrule

UniAudio 1.5 & semantic layer &  & 50 & 53 & 59 & 54 & 56                 \\
UniAudio 1.5 & semantic+acoustic layers &  & 25 & 53 & 58 & 49 & 52                 \\
\midrule
BLSP \cite{blsp} & Whisper encoder &  & 29 & 65 & 84 & 69 & 59            \\
Random  & None & & \multicolumn{3}{r}{44}  &  &  &             \\
\bottomrule
\end{tabular}
\end{table}

Speech command recognition refers to the recognition and interpretation of short phrases or keywords that are typically used to control devices or applications. In this part, we choose audio samples from the Speech Command dataset \cite{speech_command}. We choose four types of commands, including down, go, left, and right. For each command, we randomly choose 20 utterances, then we use these data to construct a 2-way-K-shot evaluation.  Experimental results are shown in Table \ref{tab:speech_command}, we can see that only using the semantic VQ layer brings the best performance. Instead, if the tokens from the acoustic layer are used, the performance will decline. One possible reason is that for the speech command recognition task, it only needs to understand the content, and the content information has been saved in the semantic layer, the additional acoustic information may disturb the LLMs's prediction. 

\subsection{Simple text-to-sound generation}
Similarly, we can also use the same setting as text-to-speech to conduct text-to-sound generation tasks. We choose a test set from the ESC50 dataset \cite{esc50}, and let the model learn to generate sound events based on the text label. For example, we can set several different sound types in the prompt, and then ask the LLAMA model to generate a new audio. However, we also find that it is hard to ask LLM to generate new types of sound. 
\begin{figure*}[t]
    \centering
    \includegraphics[width=\textwidth]{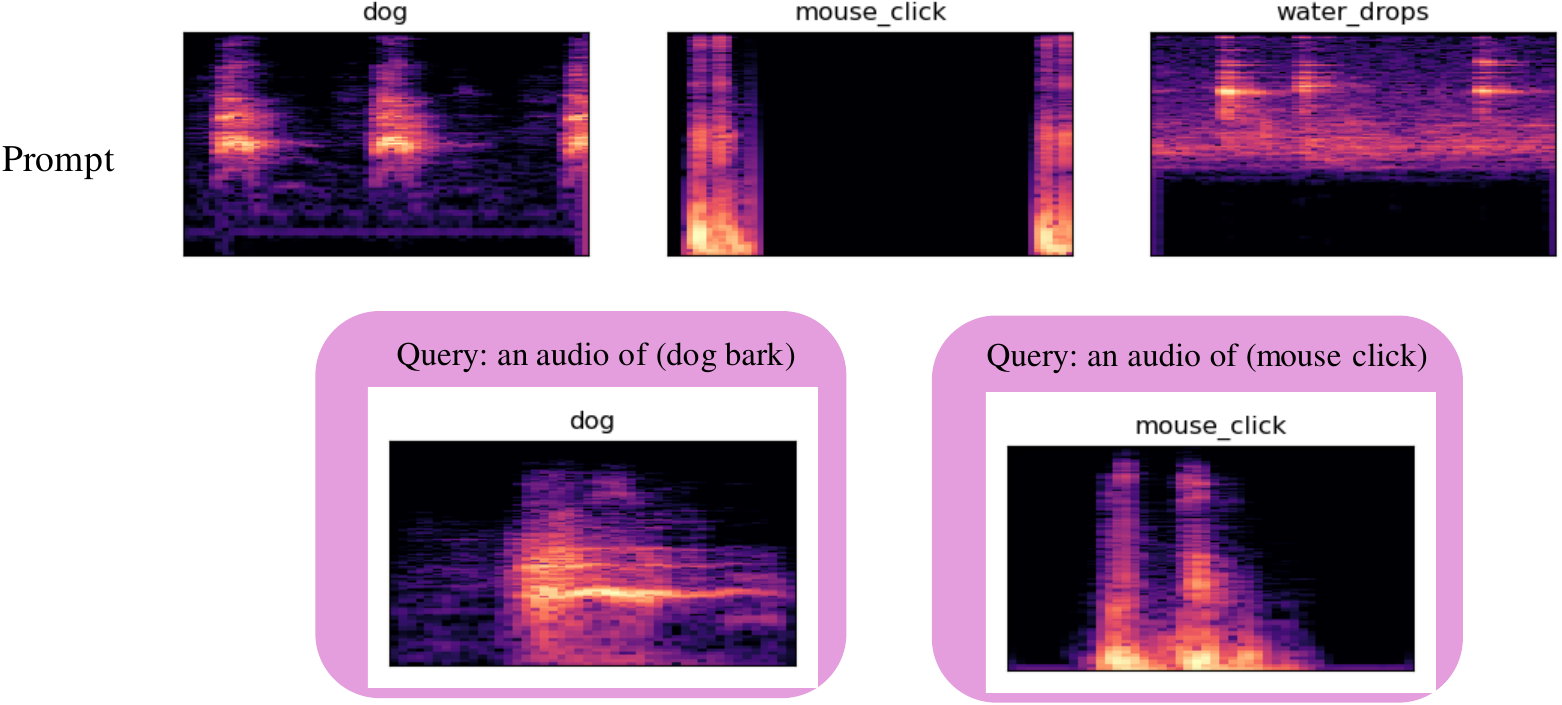}
    \caption{Examples of simple text-to-sound generation on FSDD dataset using LLM-Codec with a frozen LLAMA2 7B model.} 
    \label{fig:tta}
\end{figure*}
\subsection{Token visualization}
\begin{figure*}[t]
    \centering
    \includegraphics[width=\textwidth]{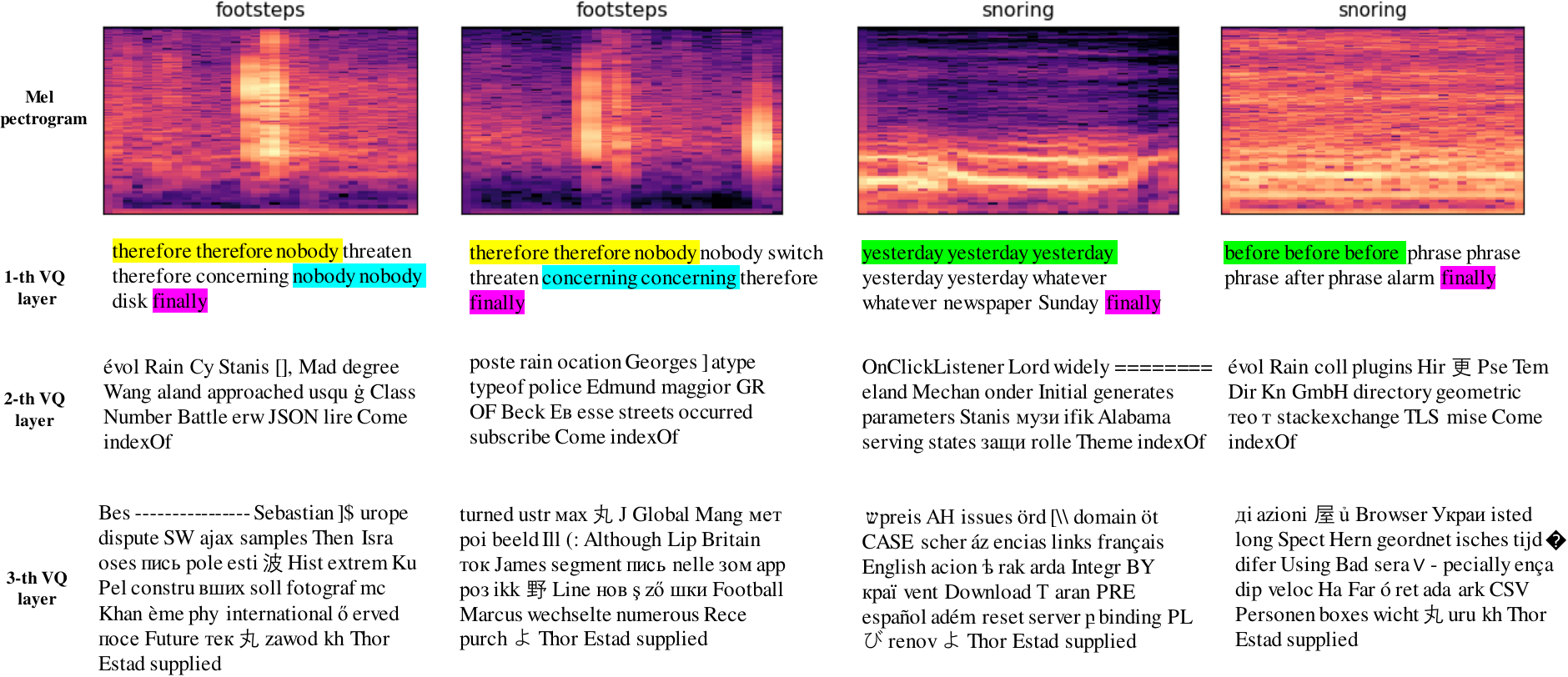}
    \caption{The token visualization of three VQ layers with LLM-Codec. The audio samples are from the ESC50 dataset.} 
    \label{fig:vis_a}
\end{figure*}
Figure \ref{fig:vis_a} shows the details of three VQ layers token visualization. We have the following findings:
(1) The few tokens in the first layers seem to more easy to understand audio's pattern. For example, we can easily find two audios that have the same sound event can be quantized into a very similar sequence. Because we force the first VQ layer to encode the semantic level information. Instead, the second and third VQ layers aims to encode the acoustic information, but these audios have obvious difference in acoustic condition (we can observe it from its mel-spectrogram). Second, it is worth noting that all of the training data is English-related, but we can see that the encoded sequence also includes other language, such as Chinese. 

\end{document}